\newcommand{\vect}[1]              
           {\mbox{\boldmath$#1$}}  
\documentclass[12pt]{article}

\begin{document}
%
\vspace*{-1.0cm}
\begin{center}
{\Large\bf Scaling in Relativistic Thomas--Fermi \\[2mm]
Approach for Nuclei}
\\[2.0cm]
S.K. Patra\footnote{Present address:
           {\it Institute of Physics, Sachivalaya Marg,
                Bhubaneswar-{\sl 751 005}, India}},
M. Centelles, X. Vi\~nas and M. Del Estal \\[2mm]
{\it Departament d'Estructura i Constituents de la Mat\`eria,
     Facultat de F\'{\i}sica,
\\
     Universitat de Barcelona,
     Diagonal {\sl 647}, E-{\sl 08028} Barcelona, Spain}
\end{center}
%
\vspace*{2.0cm}
\begin{abstract}
By using the scaling method we derive the virial theorem for the
relativistic mean field model of nuclei treated in the Thomas--Fermi
approach. The Thomas--Fermi solutions statisfy the stability condition
against scaling. We apply the formalism to study the excitation energy
of the breathing mode in finite nuclei with several relativistic
parameter sets of common use.
\end{abstract}

\mbox{}



{\it E-mail address:} mario@ecm.ub.es (M. Centelles)

\pagebreak

%

The relativistic mean field (RMF) treatment of Quantum Hadrodynamics
\cite{serot86,serot97} has proven to be very useful for describing
different properties of nuclei along the periodic table. The simplest
model, the linear $\sigma-\omega$ model of Walecka \cite{Wa74},
explains the nuclear force in terms of the exchange of $\sigma$ and
$\omega$ mesons. It is known that the value of the nuclear matter
incompressibility is unreasonably high in this linear model ($K_{\rm
nm} \sim 550$ MeV). The problem can be cured by introducing cubic and
quartic self-interactions of the $\sigma$ meson \cite{boguta77}, and
the model can be refined by adding an isovector $\rho$ meson. Current
non-linear parameter sets, such as the NL3 set \cite{lalaz97}, give
ground-state energies and densities in excellent agreement with the
experimental data, not only for magic nuclei but also for deformed
nuclei and for nuclei far from the stability line.

The scaling method has been often employed to derive the virial
theorem in the non-relativistic framework \cite{parr89}, e.g., for
nuclear effective interactions such as the Skyrme force
\cite{bohigas79}. It has also been applied in calculations of nuclear
collective excitations like the breathing mode (isoscalar giant
monopole resonance) \cite{bohigas79}. Relativistic generalizations of
the virial theorem obtained by use of the scaling method exist for
particles in external potentials \cite{brack83,lucha90}. In the RMF
model of nuclei the mean field potentials are generated
self-consistently. Owing to the meson-exchange nature of the
relativistic model one has to deal with finite range forces, which
renders the scaling more involved than for zero-range Skyrme forces.
Moreover, in contrast to the non-relativistic situation, there exist
two different densities, namely the baryon and the scalar density, in
accordance with the fact that one has two types of fields, the vector
and the scalar field.

In this letter we shall make use of the principle of scale invariance
to obtain the virial theorem for the RMF theory by working in the
Thomas--Fermi approximation. We shall include non-linear
self-couplings of the scalar field and shall deal with spherical
finite nuclei. The second derivative of the scaled energy with respect
to the scaling parameter, the so-called restoring force, turns out to
be positive (stability condition) in the Thomas--Fermi calculations.
Thus we are able to apply the method to compute the excitation energy
of the isoscalar giant monopole resonance in finite nuclei with
realistic parameter sets of the relativistic model. 

The meson field equations of the non-linear $\sigma-\omega$ model are
\cite{serot86,boguta77}
\begin{eqnarray}
(\Delta- m_{\rm s}^2)\phi & = & -g_{\rm s} \rho_{\rm s}
+b\phi^2 +c\phi^3
\label{eqFN4}  \\[1.mm]
   (\Delta - m_{\rm v}^2) V  & = &   -g_{\rm v} \rho
\label{eqFN5}  \\[1.mm]
   (\Delta - m_\rho^2)  R  & = &  - g_\rho \rho_3
\label{eqFN6}  \\[1.mm]
 \Delta {\cal A}   & = &     -e \rho_{\rm p} .
\label{eqFN7}
\end{eqnarray}
Here $\rho= \rho_{\rm p}+\rho_{\rm n}$ is the baryon density, $\rho_3=
{\textstyle{1\over2}} (\rho_{\rm p}-\rho_{\rm n})$ is the isovector
density, and $\rho_{\rm s}$ is the scalar density. The meson fields
$\phi$, $V$ and $R$ are associated with the $\sigma$, $\omega$ and
$\rho$ mesons, respectively, and $\cal A$ is the Coulomb field. It is
understood that the densities and fields are local quantities that
depend on position, even if we do not make it explicit. Units are
$\hbar= c = 1$.

Taking into account the above field equations, the relativistic energy
density of a finite nucleus in Thomas--Fermi approximation can be
written as \cite{serot86,boguta77}
\begin{equation}
{\cal H} = {\cal E} +\frac{1}{2}g_{\rm s}\phi \rho^{\rm eff}_{\rm s}
+ \frac{1}{3}b\phi^3+\frac{1}{4} c \phi^4
+\frac{1}{2} g_{\rm v}  V \rho +\frac{1}{2} g_\rho R \rho_3
+\frac{1}{2} e {\cal A} \rho_{\rm p} ,
\label{eqFN8c}\end{equation}
in terms of the nucleon energy density
\begin{equation}
{\cal E} =
\sum_{q} \frac{1}{8\pi^2} \left[k_{{\rm F}q}\epsilon^{3}_{{\rm F}q}
+k^{3}_{{\rm F}q}\epsilon_{{\rm F}q}
-{m^*}^{4}\ln\frac{k_{{\rm F}q}+\epsilon_{{\rm F}q}}{m^*}\right]
\label{eqFN2b}
\end{equation}
and of $g_{\rm s} \rho^{\rm eff}_{\rm s} = g_{\rm s} \rho_{\rm s}
-b \phi^2 - c \phi^3$, where
\begin{equation}
\rho_{\rm s} = \frac{\partial {\cal E}}{\partial m^*} =
 \sum_{q} \frac{m^*}{2\pi^2}\left[k_{{\rm F}q}\epsilon_{{\rm
F}q}-{m^*}^2 \ln\frac{k_{{\rm F}q}
+\epsilon_{{\rm F}q}} {m^*}\right]
\label{eqFN8}\end{equation}
is the scalar density and $m^* = m - g_{\rm s}\phi$ is the nucleon
effective mass. For each kind of nucleon ($q= {\rm n}, {\rm p}$) the
local Fermi momentum $k_{{\rm F}q}$ is defined by $k_{{\rm F}q}=
(3\pi^2 \rho_{q})^{1/3}$, while $\epsilon_{{\rm F}q}=\sqrt{k^2_{{\rm
F}q}+{m^*}^2}$. 

The virial theorem relates the kinetic and potential energy components
of the energy in certain circumstances. This theorem results from
homogeneity properties of the kinetic and potential energy components
of $\cal H$ with respect to a scaling transformation that preserves
the normalization. One such normalized scaled version of the baryon
density is
\begin{equation}
 \rho_\lambda (\vect{r}) = \lambda^3 \rho(\lambda\vect{r}),
\label{eqFN10}\end{equation}
where $\lambda$ is an arbitrary scaling parameter. Accordingly, the
local Fermi momentum changes as
\begin{equation}
 k_{{\rm F}q\lambda} (\vect{r}) =
 [ 3\pi^2\rho_{q\lambda}(\vect{r}) ]^{1/3}=
\lambda k_{{\rm F}q} (\lambda\vect{r}) .
\label{eqFN11}\end{equation}
The meson fields and the Coulomb field are also modified by the
scaling due to the self-consistent equations
(\ref{eqFN4})--(\ref{eqFN7}),
which will relate the scaled fields to the scaled densities.
Unfortunately, the meson fields do not scale according to simple power
laws of $\lambda$ because of the finite-range character of the meson
interactions. This is most apparent for the scalar field $\phi$, since
the scalar density in the source term of Eq.\ (\ref{eqFN4}) transforms
not only due to the scaling of $k_{{\rm F}q}$ but also of $\phi$
itself (or $m^*$), see Eq.\ (\ref{eqFN8}) for $\rho_{\rm s}$. For
reasons that will become clear immediately, we shall write the scaled
effective mass $m^*_\lambda(\vect{r})= m - g_{\rm s}
\phi_\lambda(\vect{r})$ in the form
\begin{equation}
 m^*_\lambda(\vect{r}) \equiv \lambda {\tilde m}^* (\lambda\vect{r}) .
\label{eqFN13} \end{equation}
The quantity ${\tilde m}^*$ carries an implicit dependence on
$\lambda$ apart from the parametric dependence on $\lambda\vect{r}$.

On account of Eqs.\ (\ref{eqFN11}) and (\ref{eqFN13}) the scaled form
of $\cal E$ reads
${\cal E}_\lambda(\vect{r}) =
\lambda^4 {\cal E} [ k_{{\rm F}q} (\lambda\vect{r}),
                         {\tilde m}^* (\lambda\vect{r}) ]
\equiv \lambda^4 {\tilde{\cal E}} (\lambda\vect{r})$,
while the scaled scalar density reads
$\rho_{\rm s\lambda} (\vect{r}) =
\lambda^3 \rho_{\rm s} [ k_{{\rm F}q} (\lambda\vect{r}),
                         {\tilde m}^* (\lambda\vect{r}) ]
\equiv \lambda^3 {\tilde\rho}_{\rm s} (\lambda\vect{r})$.
The tilded quantities ${\tilde{\cal E}}$ and ${\tilde\rho}_{\rm s}$
are given by Eqs.\ (\ref{eqFN2b}) and (\ref{eqFN8}) after replacing
$m^*$ by ${\tilde m}^*$. Note the usefulness of (\ref{eqFN13}) to be
able to put the transformed densities ${\cal E}_\lambda$ and
$\rho_{\rm s\lambda}$ into the above compact form. This way, for the
scaled total energy density ${\cal H_\lambda}$ we obtain
\begin{equation}
{\cal H}_\lambda  = \lambda^3 \bigg[ \lambda {\tilde{\cal E}}
+\frac{1}{2}g_{\rm s} \phi_\lambda {\tilde\rho}^{\rm eff}_{\rm s}
+\frac{1}{3} \frac{b}{\lambda^3}\phi_\lambda^3
+\frac{1}{4} \frac{c}{\lambda^3} \phi_\lambda^4
+\frac{1}{2} g_{\rm v}  V_\lambda \rho
+\frac{1}{2}g_\rho R_\lambda \rho_3
+\frac{1}{2} e {\cal A}_\lambda\rho_{\rm p} \bigg],
\label{eqFN17}\end{equation}
with the definition
$g_{\rm s}{\tilde\rho}^{\rm eff}_{\rm s} =
g_{\rm s} {\tilde\rho}_{\rm s}
-b \phi_\lambda^2/\lambda^3 - c \phi_\lambda^3/\lambda^3$.

The scaled energy is stationary for $\lambda=1$ (which leads to the
virial theorem):
\begin{eqnarray}
0 & = & \left[ \frac{\partial}{\partial \lambda}
        \int \frac{d (\lambda \vect{r})}{\lambda^3}
        {\cal H}_\lambda (\vect{r}) \right]_{\lambda=1}
\nonumber \\[1.mm]
& = &
 \int d \vect{r} \left[ {\tilde{\cal E}}
- {\tilde m}^* {\tilde\rho}_{\rm s}
-\frac{b}{\lambda^4}\phi_\lambda^3
-\frac{3}{4} \frac{c}{\lambda^4}\phi_\lambda^4 
-\frac{1}{2}g_{\rm s}{\tilde\rho}^{\rm eff}_{\rm s}
\frac{\partial \phi_\lambda}{\partial \lambda}
+\frac{1}{2}g_{\rm s} \phi_\lambda
\frac{\partial {\tilde\rho}^{\rm eff}_{\rm s}}{\partial \lambda}
\right.
\nonumber \\[1.mm]
& & \left. \mbox{}
+\frac{1}{2}g_{\rm v}\rho\frac{\partial  V_\lambda}{\partial \lambda}
+\frac{1}{2}g_\rho \rho_3 \frac{\partial R_\lambda}
{\partial \lambda}+\frac{1}{2}e\rho_{\rm p} \frac{\partial
{\cal A}_\lambda} {\partial \lambda} \right]_{\lambda=1} .
\label{eqFN19}
\end{eqnarray}
Here we have used $\partial {\tilde{\cal E}}/\partial \lambda=
{\tilde\rho}_{\rm s} \, \partial {\tilde m}^*/\partial \lambda$ (as
${\tilde\rho}_{\rm s}= \partial {\tilde{\cal E}}/\partial {\tilde
m}^*$) and, from the definition of ${\tilde m}^*$,
\begin{equation}
\frac{\partial m^*_\lambda}{\partial \lambda} =
{\tilde m}^* + \lambda \frac{\partial {\tilde m}^*}
{\partial \lambda}=-g_{\rm s} \frac{\partial \phi_\lambda}{\partial
\lambda}.
\label{eqFN20} \end{equation}

Let us exemplify the calculation of the derivatives of the scaled
fields with respect to $\lambda$ with the omega field $V_\lambda$. It
fulfils the scaled Klein--Gordon equation $(
\Delta_{\vect{\scriptstyle u}} - m_{\rm v}^2/\lambda^2 )
V_\lambda(\vect{u}) = - \lambda g_{\rm v} \rho(\vect{u})$, where we
have used Eq.\ (\ref{eqFN10}) for $\rho_\lambda$ and have switched to
the coordinate $\vect{u}= \lambda\vect{r}$. On differenciating this
equation with respect to $\lambda$ we have
\begin{equation}
\left( \Delta_{\vect{\scriptstyle u}}
- \frac{m_{\rm v}^2}{\lambda^2} \right) 
\frac{\partial V_\lambda}{\partial\lambda} =
- g_{\rm v} \rho - \frac{2 m_{\rm v}^2}{\lambda^3} V_\lambda .
\label{eqv2}\end{equation}
If one now sets $\lambda= 1$ the solution of this equation provides
$\partial V_\lambda /\partial\lambda |_{\lambda=1}$. Nevertheless, for
our purposes it is more useful to multiply both sides of (\ref{eqv2})
by $V_\lambda$, integrate over the space and then use Green's
identity on the left hand side. This way it is straightforward to get
\begin{equation}
 \frac{1}{2} \int d \vect{u} \, g_{\rm v} \rho
 \frac{\partial V_\lambda}{\partial \lambda}
= \int d \vect{u} \left[
 \frac{1}{2\lambda} g_{\rm v} \rho V_\lambda +
 \frac{1}{\lambda^4} m_{\rm v}^2 V_\lambda^2 \right] ,
\label{eqv3}\end{equation}
which at $\lambda=1$ is just one of the contributions we need in Eq.\
(\ref{eqFN19}). Analogous results are found for the rho and Coulomb
fields (with a zero mass for the latter). In the case of the scalar
field additional terms appear due to the fact that the scalar density
itself is a function of the scalar field. Following the same steps as
above, from the scaled field equation $( \Delta_{\vect{\scriptstyle
u}} - m_{\rm s}^2/\lambda^2 ) \phi_\lambda(\vect{u}) = - \lambda
g_{\rm s} {\tilde\rho}^{\rm eff}_{\rm s}(\vect{u})$ one easily arrives
at 
\begin{equation}
\left( \Delta_{\vect{\scriptstyle u}}
- \frac{m_{\rm s}^2}{\lambda^2} \right)
\frac{\partial \phi_\lambda}{\partial\lambda} =
- g_{\rm s} {\tilde\rho}^{\rm eff}_{\rm s} - \lambda g_{\rm s}
\frac{\partial {\tilde\rho}^{\rm eff}_{\rm s}}{\partial \lambda}
- \frac{2 m_{\rm s}^2}{\lambda^3} \phi_\lambda ,
\label{eqv4}\end{equation}
whence
\begin{equation}
\int d \vect{u} \left[ -\frac{1}{2} g_{\rm s} {\tilde\rho}^{\rm
eff}_{\rm s} \frac{\partial \phi_\lambda}{\partial \lambda} +
\frac{1}{2} g_{\rm s} \phi_\lambda \frac{\partial 
{\tilde\rho}^{\rm eff}_{\rm s}}{\partial \lambda} \right] =
 \int d \vect{u} \left[ - \frac{1}{2\lambda} g_{\rm s}
{\tilde\rho}^{\rm eff}_{\rm s} \phi_\lambda - \frac{1}{\lambda^4}
m_{\rm s}^2 \phi_\lambda^2 \right] .
\label{eqv5}\end{equation}

From substitution of Eqs.\ (\ref{eqv3}) and (\ref{eqv5}) (and of the
corresponding results for the rho and Coulomb fields) into Eq.\
(\ref{eqFN19}) the virial theorem for the non-linear $\sigma-\omega$
model becomes
\begin{eqnarray}
0  & = & \int d \vect{r}
\left[ {\cal E} - m^* \rho_{\rm s}
- \frac{1}{2}g_{\rm s} \phi \rho_{\rm s} - m_{\rm s}^2 \phi^2
- \frac{1}{2}b\phi^3 - \frac{1}{4} c \phi^4
+\frac{1}{2} g_{\rm v}  V \rho + m_{\rm v}^2 V^2 \right.
\nonumber \\[1.mm]
& & \left. \mbox{}
+\frac{1}{2} g_\rho R \rho_3 + m_\rho^2 R^2
+\frac{1}{2} e {\cal A} \rho_{\rm p} \right] .
\label{eqFN21d}\end{eqnarray}
Actually, introducing the kinetic energy density $\tau$ we have ${\cal
E} - m^* \rho_{\rm s} = \tau + m \rho- m \rho_{\rm s}$, which makes
more obvious the kinetic energy component in the virial theorem.
Using Eq.\ (\ref{eqFN21d}) to eliminate $\cal E$ from the expression
of the relativistic energy density $\cal H$, the RMF energy of a
nucleus takes the remarkably simple form
\begin{equation}
\int d \vect{r}  [ {\cal H} - m \rho ] =
\int d \vect{r}  \left[
m (\rho_{\rm s} - \rho) + m_{\rm s}^2 \phi^2
- m_{\rm v}^2 V^2 - m_\rho^2 R^2 + \frac{1}{3} b\phi^3 \right] ,
\label{eqFN21h}\end{equation}
where we have subtracted the nucleon rest mass contribution. This
expression shows very clearly the relativistic mechanism for nuclear
binding. It stems from the cancellation between the scalar and vector
potentials and from the difference between the scalar and the baryon
density (i.e., from the small components of the wave functions).
Equations (\ref{eqFN21d}) and (\ref{eqFN21h}) are satisfied not only
by the Thomas--Fermi solutions, but also by the ground-state densities
and meson fields obtained from a quantal Hartree calculation. Of
course, the energy stationarity condition against dilation of the RMF
problem must be fulfilled by any approximation scheme utilized to
solve it.

As a further application of the method we shall use it in calculations
of the isoscalar giant monopole resonance (ISGMR). It is customary to
write the excitation energy of the ISGMR as
\begin{equation}
 E_{\rm M} = \sqrt{ \frac{C_{\rm M}}{B_{\rm M}} } ,
\label{eqFN36b}
\end{equation}
where $C_{\rm M}$ and $B_{\rm M}$ are called, respectively, the
restoring force (or incompressibility of the finite nucleus) and the
mass parameter of the monopole vibration. To study $E_{\rm M}$ in the
RMF the authors of Refs.\ \cite{nishizaki87,zhu91} resorted to a local
Lorentz boost and the scaling method. Following these works one has
\begin{equation}
 C_{\rm M} =  \frac{1}{A} \left[ \frac{\partial^2}{\partial \lambda^2}
\int  d \vect{r} {\cal H}_\lambda (\vect{r}) \right]_{\lambda=1} ,
\label{eqFN36}\end{equation}
where the scaling parameter $\lambda$ now plays the role of the
collective coordinate of the monopole vibration, and
\begin{equation}
B_{\rm M} = \frac{1}{A} \int d\vect{r} r^2 {\cal H}(\vect{r}) ,
\label{eqFN37}\end{equation}
with $A$ being the mass number of the nucleus. The investigations of
Refs.\ \cite{nishizaki87,zhu91} were restricted to the linear
$\sigma-\omega$ model, either for nuclear matter \cite{nishizaki87} or
for symmetric and uncharged finite nuclei (with the densities solved
in Thomas--Fermi approximation) \cite{zhu91}. It is well known that
the surface properties of nuclei cannot be described within the linear
model, and hence nor can the overall properties of nuclei.

To compute $C_{\rm M}$ it is easiest to replace the relations
(\ref{eqv3}) and (\ref{eqv5}) into the expression of $\partial [ \int
d \vect{r} {\cal H}_\lambda (\vect{r})] / \partial\lambda$ and derive
again with respect to $\lambda$. After some algebra we obtain the
restoring force as
\begin{eqnarray} 
 C_{\rm M} & = &
\frac{1}{A} \int d \vect{r} \left[
- m \frac{\partial {\tilde\rho}_{\rm s}}{\partial \lambda} 
+ 3 \left( m_{\rm s}^2 \phi^2 + \frac{1}{3} b \phi^3 
- m_{\rm v}^2 V^2 - m_\rho^2 R^2 \right)  \right.
\nonumber \\[1.mm]
& & \left. \mbox{}
- (2 m_{\rm s}^2 \phi + b\phi^2) 
\frac{\partial \phi_\lambda}{\partial\lambda}
+ 2 m_{\rm v}^2 V
\frac{\partial V_\lambda}{\partial\lambda} 
+ 2 m_\rho^2 R
\frac{\partial R_\lambda}{\partial\lambda} \right]_{\lambda=1} ,
\label{eqFN30}\end{eqnarray}
where
\begin{equation}
\left.  \frac{\partial{\tilde\rho}_{\rm s}}
{\partial \lambda} \right|_{\lambda=1}  =
\left. \frac{\partial{\tilde\rho}_{\rm s}}{\partial {\tilde m}^*}
\frac{\partial {\tilde m}^*}{\partial \lambda} \right|_{\lambda=1}  =
 - \frac{\partial \rho_{\rm s}}{\partial m^*}
\left[ m^* + g_{\rm s} \frac{\partial \phi_\lambda}
{\partial \lambda} \right]_{\lambda=1} .
\label{eqFN245}\end{equation}
The derivatives of the scaled meson fields with respect to $\lambda$
are computed by solving Eqs.\ (\ref{eqv2}) and (\ref{eqv4}) at
$\lambda=1$. We have found $C_{\rm M}$ to be positive for all of the
(linear and non-linear) parameter sets we have tested in the
Thomas--Fermi calculations. A large part of the final value of $C_{\rm
M}$ (usually far more than a half) is due to the contribution of the
term $\partial {\tilde\rho}_{\rm s}/\partial \lambda|_{\lambda=1}$.

In Table 1 we display the calculated Thomas--Fermi excitation energies
of the ISGMR, together with the empirical estimate $E_{\rm M} \sim
80/A^{1/3}$ MeV \cite{woude87}, for $^{40}$Ca, $^{90}$Zr, $^{116}$Sn,
$^{144}$Sm and $^{208}$Pb. Recent experimental data on the centroid
energy of the ISGMR are available for these nuclei \cite{young99}. We
have employed the non-linear parameter sets NL-Z2 ($K_{\rm nm}=172$
MeV) \cite{bender99}, NL1 ($K_{\rm nm}=212$ MeV) \cite{reinhard86},
NL3 ($K_{\rm nm}=272$ MeV) \cite{lalaz97}, NL-SH ($K_{\rm nm}=355$
MeV) \cite{sharma93} and NL2 ($K_{\rm nm}=399$ MeV) \cite{lee86}.
These parameter sets have been determined by least-squares fits to
ground-state properties of a few spherical nuclei and are of common
use in RMF calculations. From the table one can see that the smaller
the mass number, the larger is the monopole energy. The energy of the
ISGMR increases with increasing $K_{\rm nm}$ in the various parameter
sets. For example, the monopole energy in $^{208}$Pb is 12.3 MeV for
NL-Z2 while it is 18.1 MeV for NL2. The dependence on $K_{\rm nm}$ is
roughly linear for each nucleus.

In assuming nuclear matter within a certain volume the authors of
Ref.\ \cite{nishizaki87} estimated the monopole excitation energy of a
finite nucleus as
\begin{equation}
E_{\rm M} =
\sqrt{ \frac{K_{\rm nm}}{ \langle r^2 \rangle 
(\epsilon_{\rm Fnm} + g_{\rm v}  V_{\rm nm})} } ,
\label{eqFN55}\end{equation}
with $\langle r^2 \rangle= \frac{3}{5} R^2$ and $R= 1.2 A^{1/3}$ fm.
They evaluated (\ref{eqFN55}) for the linear model of Walecka and
found
$E_{\rm M}=160/A^{1/3}$ MeV, which has the correct dependence on the
mass number but is twice as large as the empirical value. From Eq.\
(\ref{eqFN55}) one finds $E_{\rm M}= 92$, 102, 115, 132 and
$140/A^{1/3}$
MeV for the non-linear sets NL-Z2, NL1, NL3, NL-SH and NL2,
respectively. Comparing with Table 1, the finite size effects reduce
the prediction obtained from nuclear matter by a factor ranging from
$\sim 1.4$ in $^{40}$Ca to $\sim 1.3$ in $^{208}$Pb, rather
independently of the parameter set.

The ISGMR has been studied in the time-dependent RMF (TDRMF) theory by
Vretenar et al.\ \cite{vretenar97}. We include in Table 1 their
results for the energy of the main peaks that appear in the monopole
strength distributions of $^{90}$Zr and $^{208}$Pb. Our scaling
results compare very well in the case of $^{208}$Pb for all parameter
sets, but give somewhat larger excitation energies for $^{90}$Zr. It
should be mentioned that the Fourier spectrum of $^{90}$Zr in the
TDRMF calculation is considerably fragmented (specially for the sets
with higher $K_{\rm nm}$) and then the determination of the centroid
energy is more uncertain \cite{vretenar97}. Very recently, it has been
demonstrated that the relativistic random phase approximation (RRPA)
is equivalent to the small amplitude limit of the TDRMF theory in the
no-sea approximation, when pairs formed from the empty Dirac sea
states and the occupied Fermi sea states are included in the RRPA
\cite{ma01}. 

Microscopic calculations of ISGMR energies in nuclei are a valuable
source of information on the nuclear compression modulus $K_{\rm nm}$
\cite{blaizot95,farine97}, which is an important ingredient not only
for finite nuclei but also for heavy ion collisions, supernovae and
neutron stars. A further inspection of Table 1 shows that the
empirical law $E_{\rm M} \sim 80/A^{1/3}$ MeV lies between the
predictions of the NL1 and NL3 sets, as expected from the reasonable
value of $K_{\rm nm}$ in these interactions. On the contrary, $K_{\rm
nm}$ is too high in NL-SH and NL2 and we see that these sets
overestimate the empirical curve and the experimental data for all
nuclei of Table~1. No RMF parameter set seems capable of reproducing
the mass-number dependence of the experimental data over the whole
analyzed region, particularly in the lighter nuclei. One should note,
however, that our calculation provides a prediction for the mean value
or centroid of the excitation energy of the resonance. To establish a
link between $K_{\rm nm}$ and the measured energies the most
favourable situation is then met in heavy nuclei, where the
experimental strength is less fragmented than in medium and light
nuclei \cite{young99}. If we only take into account the data of
$^{144}$Sm and $^{208}$Pb, our results of Table 1 suggest that $K_{\rm
nm}$ of a RMF interaction should belong to the range 225--255 MeV\@.
(If we disregard the set NL-Z2, as in Refs.\ \cite{vretenar97,ma01},
the range is 230--260 MeV\@.) From their TDRMF and RRPA calculations
the authors of Refs.\ \cite{vretenar97,ma01} conclude that the value
of $K_{\rm nm}$ should be close to 250--270 MeV\@. Non-relativistic
Hartree--Fock plus RPA analyses using Skyrme and Gogny interactions
determine $K_{\rm nm}$ to be $215\pm15$ MeV \cite{blaizot95,farine97},
thus lower than in the RMF model.

We have derived the virial theorem for the relativistic nuclear mean
field model on the basis of the scaling method and the Thomas--Fermi
approximation. In this approach we have calculated for realistic
parameter sets of the RMF theory the breathing-mode energy of finite
nuclei fully self-consistently (i.e., we did not use a leptodermous
expansion of the finite nucleus incompressibility as in some previous
studies with the scaling method \cite{stoitsov94}). The present
calculations extend earlier work performed with the linear
$\sigma-\omega$ model \cite{nishizaki87,zhu91}.

The excitation energies of the monopole oscillation turn out to be in
good agreement with the outcome of dynamical time-dependent RMF
calculations. It has been shown very recently that the relativisitc
RPA, with the inclusion of Dirac sea states, amounts to the limit of
small amplitude oscillations of the TDRMF theory \cite{ma01}. From the
present Thomas--Fermi analysis one can thus conclude that, similarly
to the non-relativistic case, also in the relativistic framework the
excitation energies obtained with the scaling method simulate the
results of the random phase approximation.

%
\section*{Acknowledgements}
We thank J. Navarro and Nguyen Van Giai for valuable discussions.
Support from the DGICYT (Spain) under grant PB98-1247 and from DGR
(Catalonia) under grant 2000SGR-00024 is acknowledged. S.K.P. thanks
the Spanish grant SB97-OL174874 for financial support.

\pagebreak
%

%
\pagebreak

%
\begin{table}
\caption{Excitation energy of the monopole state (in MeV) obtained in
the scaling approach by using various relativistic parameter sets (in
order of increasing value of the compression modulus $K_{\rm nm}$).
The energies of the main peaks found in the time-dependent RMF
calculations of Ref.\ \cite{vretenar97} are also shown for $^{90}$Zr
and $^{208}$Pb. The experimental centroid energies are from Ref.\
\cite{young99}.}
\vspace{0.5cm}
\begin{center}
\begin{tabular}{rlclclclclclclc}
\hline
  && NL-Z2 && NL1 && NL3 && NL-SH && NL2 && $80A^{-1/3}$ && Exp.\ \\
\hline
$^{40}$Ca & &20.5 & &21.2 & & 23.5 & & 26.6& & 29.5 & & 23.4 &&
 $19.2\pm0.4$\\
$^{90}$Zr & &16.4 & &17.2 & & 19.2 & & 21.9& & 24.0 & & 17.9 &&
 $17.9\pm0.2$\\
\cite{vretenar97}
 & & & & 15.7 & & $\sim 18$ & &   & &  & &  && \\
$^{116}$Sn& &15.1 & &15.9 & & 17.7 & & 20.3& & 22.3 & & 16.4 &&
 $16.1\pm0.1$\\
$^{144}$Sm& &14.1 & &14.9 & & 16.6 & & 19.0& & 20.8 & & 15.3 &&
 $15.4\pm0.3$\\
$^{208}$Pb& &12.3 & &12.9 & & 14.5 & & 16.6& & 18.1 & & 13.5 &&
 $14.2\pm0.3$\\
\cite{vretenar97}
& & & & 12.4 & & 14.1 & & 16.1& & 17.8 & &  && \\
\hline
\end{tabular}
\end{center}
\end{table}


\begin{thebibliography}{99}
%
\parskip= -1.5mm
%
\bibitem{serot86} B.D. Serot and J.D. Walecka,
                  Adv.\ Nucl.\ Phys.\ {\bf 16} (1986) 1.
\bibitem{serot97} B.D. Serot and J.D. Walecka, Int.\ J.
                  of Mod.\ Phys.\  {\bf E6} (1997) 515.
\bibitem{Wa74}  J.D. Walecka,
                Ann.\ Phys.\ (N.Y.) {\bf 83}, 491 (1974).
\bibitem{boguta77}  J. Boguta and A.R. Bodmer, Nucl.\ Phys.\ {\bf
                    A292} (1977) 413.
\bibitem{lalaz97} G.A. Lalazissis, J. K\"oning and P. Ring,
              Phys.\ Rev.\ {\bf C55} (1997) 540.
%
\bibitem{parr89} For scaling relations and the virial theorem see,
e.g., R.G. Parr and W. Yang, Density-Functional Theory of Atoms
and Molecules (Oxford University Press, New York, 1989), and
references therein.
%
\bibitem{bohigas79} O. Bohigas, A. Lane and J. Martorell,
                    Phys.\ Rep.\ {\bf 51} (1979) 267.
%
\bibitem{brack83} M. Brack, Phys.\ Rev.\ {\bf D27} (1983) 1950.
%
\bibitem{lucha90} W. Lucha and F.F. Sch\"oberl,
                  Phys.\ Rev.\ Lett.\ {\bf 64} (1990) 2733.
\bibitem{nishizaki87} S. Nishizaki, H. Kurasawa and T. Suzuki,
                      Nucl.\ Phys.\ {\bf A462} (1987) 689.
\bibitem{zhu91} Chaoyuan Zhu and Xi-Jun Qiu,
                J. Phys.\ {\bf G17} (1991) L11.
\bibitem{woude87} A. van der Woude,
                Prog.\ Part.\ Nucl.\ Phys.\ {\bf 18} (1987) 217.
\bibitem{young99} D.H. Youngblood, H.L. Clark and Y.-W. Lui,
                   Phys.\ Rev.\ Lett.\ {\bf 82} (1999) 691;
                  D.H. Youngblood, Y.-W. Lui and H.L. Clark,
                   Phys.\ Rev.\ {\bf C63} (2001) 067301.
\bibitem{bender99} M. Bender, K. Rutz, P.-G. Reinhard, J.A. Maruhn
                   and W. Greiner,
                   Phys.\ Rev.\ {\bf C60} (1999) 034304.
\bibitem{reinhard86} P.-G. Reinhard, M. Rufa, J. Maruhn, W. Greiner
                     and
                     J. Friedrich, Z. Phys.\ {\bf A323} (1986) 13.
\bibitem{sharma93} M.M. Sharma, M.A. Nagarajan and P. Ring,
               Phys.\ Lett.\ {\bf B312} (1993) 377.
\bibitem{lee86} S.J. Lee, J. Fink, A.B. Balantekin, M.R. Strayer, A.S.
       Umar,   P.-G. Reinhard, J.A. Maruhn and  W. Greiner,
               Phys.\ Rev.\ Lett.\ {\bf 57} (1986) 2916.
\bibitem{vretenar97} D. Vretenar, G.A. Lalazissis,
                     R. Behnsch, W. P\"oschl and
                     P. Ring,  Nucl.\ Phys.\ {\bf A621} (1997) 853.
\bibitem{ma01} Zhong-yu Ma, Nguyen Van Giai, A. Wandelt, D. Vretenar
               and P. Ring, Nucl.\ Phys.\ {\bf A686} (2001) 173;
       P. Ring, Zhong-yu Ma, Nguyen Van Giai, D. Vretenar, A. Wandelt
       and Li-gang Cao, Nucl.\ Phys.\ {\bf} (2001), in press.
\bibitem{blaizot95} J.P. Blaizot, J.F. Berger, J. Decharg\'e and
                    M. Girod, Nucl.\ Phys.\ {\bf A591} (1995) 435.
\bibitem{farine97} M. Farine, J.M. Pearson and F. Tondeur,
                   Nucl.\ Phys.\ {\bf A615} (1997) 135.
\bibitem{stoitsov94} D. Von-Eiff, J.M. Pearson, W. Stocker and M.K.
                     Weigel, Phys.\ Rev.\ {\bf C50} (1994), 831;
                     M.V. Stoitsov, M.L. Cescato, P. Ring and M.M.
                     Sharma, J. Phys.\ {\bf G20} (1994) L149;
                     T. v.\ Chossy and W Stocker,
                     Phys.\ Rev.\ {\bf C56} (1997) 2518.
%
\end{thebibliography}
\end{document}